\documentclass[aps,twocolumn,pra,superscriptaddress,amsmath,showpacs,tightenlines]{revtex4-1}
\usepackage{amssymb}
\usepackage{amsmath}
\usepackage{graphicx}
\usepackage{subfigure}
\usepackage{natbib}
\usepackage{epsfig}
\usepackage{amsfonts}
\usepackage{mathrsfs}
\usepackage{xcolor}
\usepackage[toc,page,title,titletoc,header]{appendix}
\begin{document}

\title{Single-photon scattering and bound states in a one-dimensional waveguide with topological giant atom}

\author{Wei Zhao}
\affiliation{Center for Quantum Sciences and School of Physics, Northeast Normal University, Changchun 130024, China}
\author{Tian Tian}
\affiliation{School of Materials Science and Engineering, Changchun University, Changchun 130022, China}
\author{Zhihai Wang}
\email{wangzh761@nenu.edu.cn}
\affiliation{Center for Quantum Sciences and School of Physics, Northeast Normal University, Changchun 130024, China}

\begin{abstract}
We investigate the single photon scattering and bound states in a coupled resonator waveguide (CRW) which couples to a topological giant atom (TGA) via two distant sites. Here, the TGA is constructed by a one dimensional Su-Schrieffer-Heeger (SSH) chain with finite length. By modulating the topological phase of the TGA, the incident photon in the CRW can be completely reflected or transmitted, and is therefore beneficial to design the coherent photonic device. Meanwhile, we also achieve two pairs of bound states locating respectively above and blow the continuum. Whether the gap is open or closed depends on the boundary condition of the TGA. Therefore, the combination of the topology and the interference provides us an exciting opportunity to manipulate the photonic state in the context of waveguide quantum electrodynamics.
\end{abstract}

\date{\today}

\maketitle

\section{INTRODUCTION}

The light-matter interaction plays a crucial role in the fundamental sciences~\cite{forp2019,gutr2021},  underpinning the rapid development of quantum technology. Recently, the light-matter interaction in waveguide structures has attracted a lot of attentions and brought a lot of theoretical~\cite{zhay2021} and experimental studies in the waveguide quantum electrodynamics (QED)~\cite{gux2017} community, such as dressed or bound states~\cite{zheh2010,calg2016,shit2016,sane2017,fonp2017,calg2019,zhaw2020}, phase transitions~\cite{fitm2017,qial2019}, single-photon devices~\cite{shej2005,chad2007,zhol2008}, exotic topological and chiral phenomena~\cite{rinm2014,yanv2014,miri2016,gong2016,belm2019,mahs2020}. At the same time, the dynamical control of single-photon transmission is a hot topic in constructing quantum networks~\cite{kimh2008}. Since photons provide a reliable output of quantum information and the single-photon is considered to be one of the most suitable carriers of quantum information. In quantum devices and quantum networks~\cite{lodp2017,royd2017,chad2018}, one-dimensional (1D) waveguides~\cite{rits2012,lodp2018} are essential light-matter interfaces, and the controllable single-photon transport with linear and nonlinear dispersion relationships has been extensively studied~\cite{caiq2021,zhangxj2023,lix2023}.

In the traditional treatment, the light-matter interaction is often modeled by the dipole approximation, where the atoms are treated as point-like dipoles~\cite{wald2008}. However, the study of the interaction in atom-waveguide systems has been extended to the interaction between the photon and giant atom(s) recently, where the non-local light-matter interaction occurs with multiple points.  In the giant atom community~\cite{pett2003,koca2014,koca2018,turp2019,andg2019,tuda2019,guos2020,Kanb2020,bagr2023}, since the interactions between the waveguide photons and the giant atoms include multi-path quantum interferences, many new phenomena those do not exist in the traditional small atom systems have been predicted, such as tunable bound state~\cite{guol2020,lons2021,qiuy2023,limk2023}, decoherence free interaction~\cite{cara2020}, electromagnetically induced transparency~\cite{abda2022,vada2021}, Autler-Townes splitting phenomena~\cite{zhaw2022}, chiral physics~\cite{wanx2021,wanx2022,liun2022,Wagx2022,sora2022}, and many others~\cite{dul2021,Dul2021,yuh2021}. The physical basis behind these phenomena is the interference and delay effects in the propagation of the photon/phonon between different coupling points.

On the other hand, Haldane and Raghu~\cite{half2008,rags2008} proposed to manipulate the photon transport via topological structure in 2008, which paved the way for the development of topological photonics~\cite{blik2015,Blik2015,blik2019,ozat2019,jang1,jang2,jang3,jang4}. In the simplest one dimensional case, the topological waveguide can be constructed~\cite{gux,caiw2019} through the Su-Schrieffer-Heeger (SSH)~\cite{suw1979} chain, which is characterized by nonzero winding number or zero-mode edge state(s) in the topological nontrivial phase with periodical boundary condition (PBC) or open boundary condition (OBC), respectively~\cite{xuhs2022,jinl2017,jinl2019,wuhc2021}. {Especially, the edge state in one dimensional SSH chain has been experimentally observed, and the localization length has been extracted in terms of the survival probability in Ref. \cite{wangl2022}.}

Therefore, combining the unique effect of the nonlocal light matter interaction in giant atom quantum optics and the robust natural of the topological photonics, we here investigate the manipulation of single photon state in the context of waveguide QED. To this end, we couple a one dimensional SSH chain with finite length to an infinite CRW, to study the photonic scattering and dressing state in the single excitation level. Since the short topological SSH chain supports a few discrete energy levels and couples to the CRW via two distant site, we name it as topological giant atom (TGA) in what follows. Thus, the topological nature of the TGA is utilized to adjust the photon state in the CRW. Our findings imply that the single photon scattering can be modulated by the topological phase of the TGA and the bound state is controlled by its boundary conditions.

\section{MODEL AND HAMILTONIAN}
\label{model}

\begin{figure}
\centering
\includegraphics[width=\linewidth]{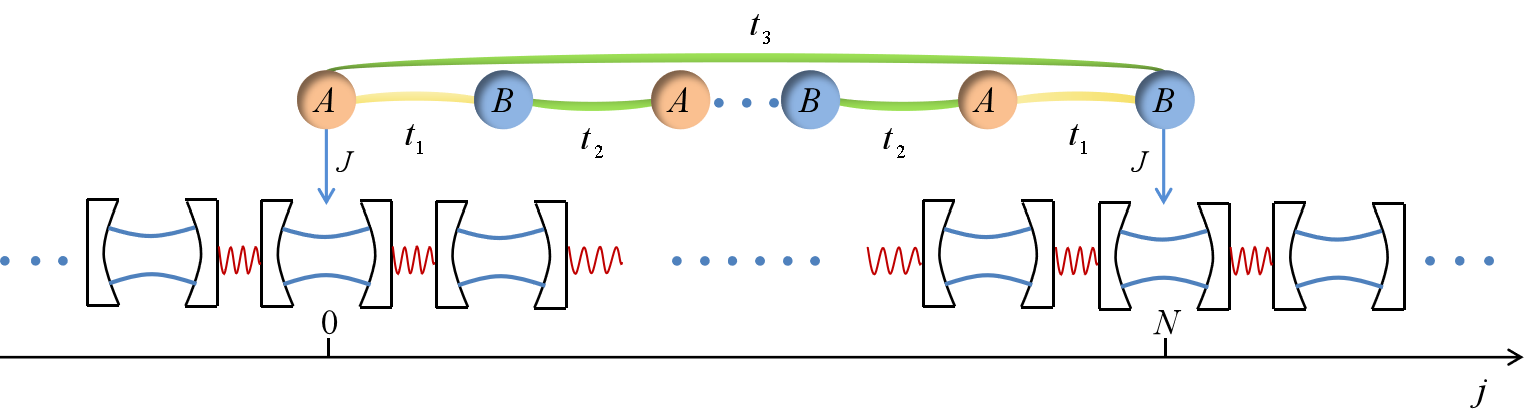}
\caption{Schematic diagram of coupling of a TGA to an infinite CRW via two coupling points. The upper part is a TGA formed by the finite SSH chain with $t_1$ and $t_2$ being intra- and inter-cell hopping amplitudes, respectively, and $t_3$ is the cyclic hopping amplitude. The lower part is an infinite CRW. The TGA coupled to the CRW via two separate sites of $j=0$ and $j=N$, and the coupling strengths for the two sites are both $J$.}
\label{f1}
\end{figure}

As schematically shown in Fig.~\ref{f1},  the system we consider is composed of two arrays of CRW. The upper one is a $N+1$ sites [i.e., $(N+1)/2$ pairs for odd $N$] SSH chain and the lower one is an infinite CRW with uniform photonic hopping strength. Since the upper SSH chain supports a finite of discrete energy levels and serves as TGA, whose topological nature is discussed below. The TGA couples to the CRW non-locally via two separate sites of $j=0$ and $j=N$. The Hamiltonian $H$ of the system can be divided into three parts, i.e., $H=H_{c}+H_{S}+H_{I}$.

The first part $H_{c}$ represents the free Hamiltonian of the CRW, and is expressed as
\begin{equation}
\label{eq:1}
H_{c}=\omega _c\sum_{j}a_{j}^{\dagger}a_{j}-\xi\sum_{j=-\infty}^{+\infty}\left(a_{j+1}^{\dagger}a_{j}+a_{j}^{\dagger}a_{j+1}\right),
\end{equation}
where $\omega_c$ is bare frequency of the resonators and $a_j^{\dagger}$ ($a_j$) is the bosonic creation (annihilation) operator on site $j$, and satisfies the commutation relation $[a_{j},a_{j}^{\dagger}]=1$. $\xi$ is the hopping strength between the nearest resonators.

The second part $H_{S}$ of the Hamiltonian $H$ represents the free Hamiltonian of the TGA [consisting of $L$ unit cells, $L=\left(N+1\right)/2$ for odd $N$], which can be written as
\begin{equation}
\label{eq:2}
\begin{split}
H_{S}&=\sum_{l=1}^{L}\omega_e\left(C_{A,l}^{\dagger}C_{A,l}+C_{B,l}^{\dagger}C_{B,l}\right)\\
&+t_{1}\sum_{l=1}^{L}\left(C_{A,l}^{\dagger}C_{B,l}+{\rm H.c.}\right)\\
&+t_{2}\sum_{l=1}^{L-1}\left(C_{A,l+1}^{\dagger}C_{B,l}+{\rm H.c.}\right)\\
&+t_{3}\left(C_{A,1}^{\dagger}C_{B,L}+{\rm H.c.}\right),
\end{split}
\end{equation}
In the TGA, each unit cell hosts $A$ and $B$ two resonators with the identical frequency $\omega_e$, and $C_{A,l}\left(C_{B,l}\right)$ is the bosonic annihilation operator at the $A(B)$ sublattices of the $l$th unit cell (see Fig.~\ref{f1}). The intra- and inter-cell hopping amplitudes are $t_{1}$ and $t_{2}$, respectively. Here, we use the value of $t_3$ to distinguish the boundary condition of the TGA.
For the PBC, we set $t_{3}=t_{2}$. {In the case of PBC, the topological phase transition is characterized by the winding number $w$. In the topological trivial phase ($|t_1|>|t_2|$), the winding phase is $w=1$.
Otherwise, in the the topological nontrivial phase ($|t_1|<|t_2|$), the winding phase is $w=0$.} For the OBC, we set $t_{3}=0$, and the TGA supports the boundary state in the topologically nontrivial phase, but not in the trivial phase.

The third part $H_{I}$ of the Hamiltonian describes the coupling between the TGA and the CRW via the $0$th and $N$th resonators with the same coupling strength $J$. Under the rotating wave approximation, the Hamiltonian $H_{I}$ can be written as
\begin{equation}
\label{eq:3}
H_{I}=J\left(a_{0}^{\dagger}C_{A,1}+a_{N}^{\dagger}C_{B,L}+{\rm H.c.}\right).
\end{equation}

\section{SINGLE-PHOTON SCATTERING }
\label{scattering}

In this section, we will discuss the behavior of the single-photon scattering by an open TGA  ($t_3=0$). We consider that a single-photon with wave vector $k$ is incident from the left side of the waveguide.
Since the excitation number in the system is conserved, the eigenstate in the single-excitation subspace can be written as
\begin{equation}
\label{eq:4}
\left|E_{k}\right\rangle =\sum_{j}U_{j}a_{j}^{\dagger}\left|G\right\rangle +\sum_{l}X_{l}C_{A,l}^{\dagger}\left|G\right\rangle +\sum_{l}Y_{l}C_{B,l}^{\dagger}\left|G\right\rangle,
\end{equation}
where $\left |G  \right \rangle$ represents that all of the resonators in the CRW and TGA are in their vacuum states. $U_j$ is the probability amplitude for finding a photonic excitation in site $j$ of the CRW , and $X_{l}(Y_{l})$ describe the excitation amplitudes in site $A_l(B_l)$ of the TGA.
In the regimes of $j<0$, $0\leqslant j\leqslant N$ and $j>N$, the amplitude $U_j$ can be written in the form
\begin{equation}
\label{eq:5}
\begin{split}
U_j=\begin{cases}e^{ikj}+re^{-ikj},&j<0\\Ae^{ikj}+Be^{-ikj},& 0\leqslant j\leqslant N
\\te^{ikj},&j>N,
\end{cases}
\end{split}
\end{equation}
where $r$ and $t$ are respectively the single-photon reflection and transmission amplitudes. Furthermore, the second line in the above equation implies that the incident photon can be transmitted (reflected) by the left (right) leg of the TGA with amplitude $A$ ($B$). The transmitted photon then propagates back and forth in the spatial regime covered by the TGA with $0\leqslant j\leqslant N$.

Solving the Schr\"{o}dinger equation $H\left|E_{k}\right\rangle=E\left|E_{k}\right\rangle$ in the region of $j\neq0,N$, it yields a dispersion relationship of $E_k=\omega _c-2\xi\cos k$. Furthermore, the continuity conditions at $j=0$ and $j=N$ give $1+r=A+B$ and $Ae^{ikN}+Be^{-ikN}=te^{ikN}$, respectively.

\begin{figure}[ht]
\centering
\includegraphics[width=0.90\columnwidth]{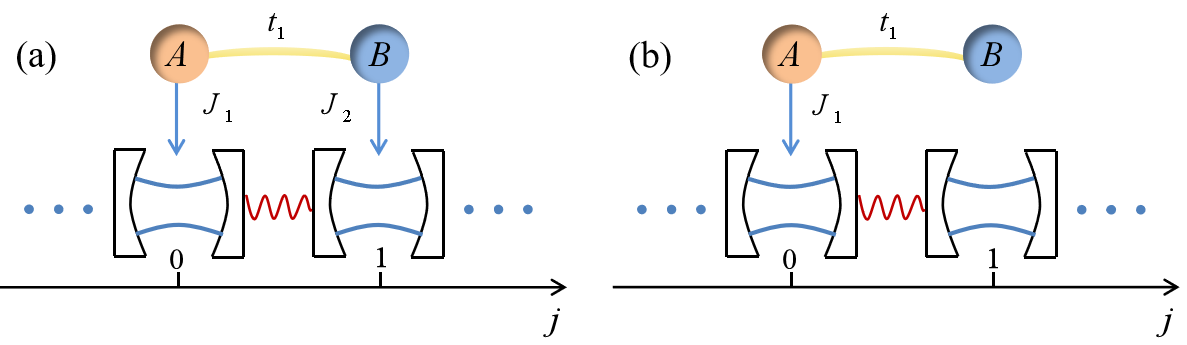}
\includegraphics[width=0.89\columnwidth]{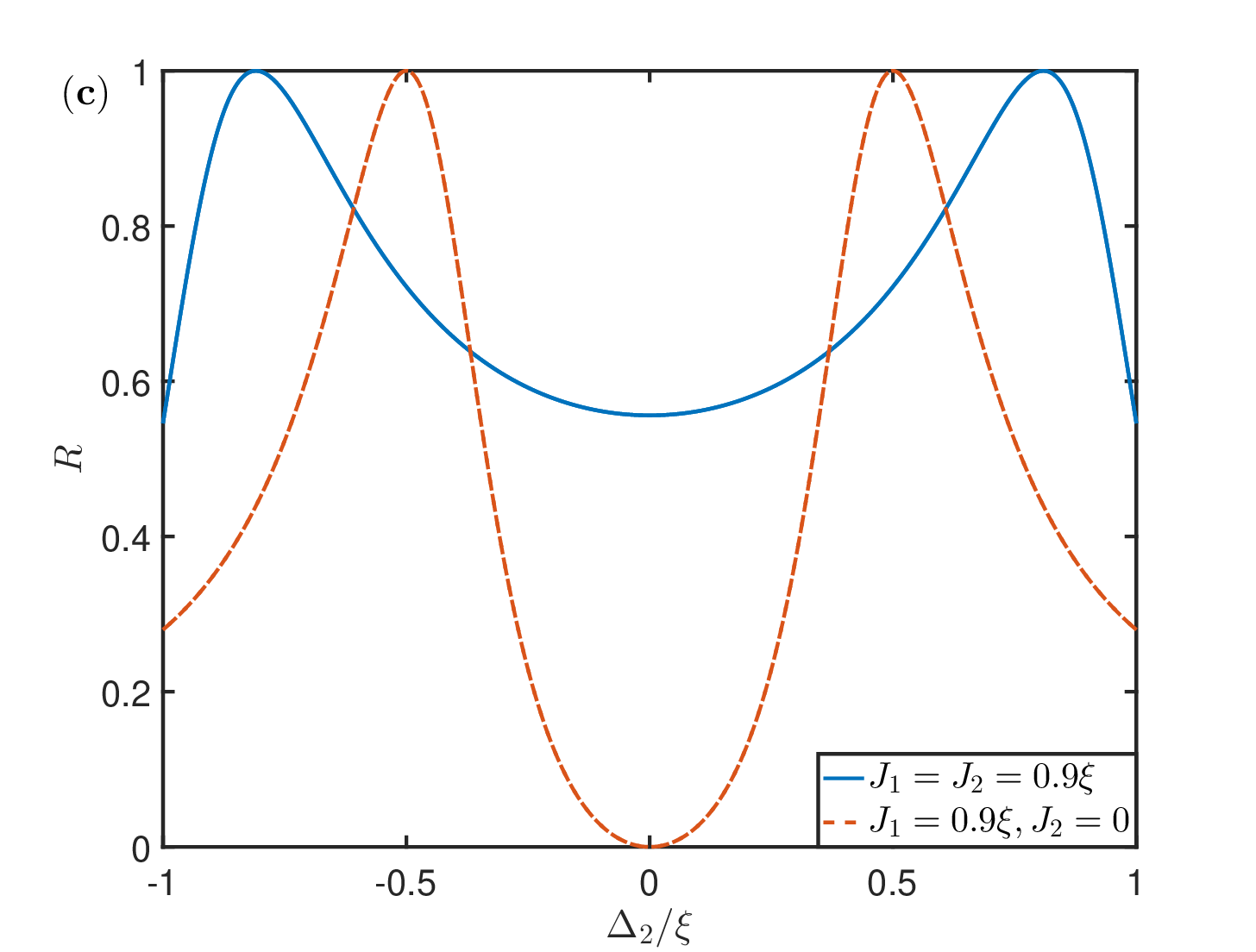}
\caption{(a) Schematic diagram of the model when $N=1$. (b) Schematic diagram of TGA where only one of the two sites is couple to CRW via one resonator. (c) The reflection rate $R$ as a function of detuning $\Delta_2$ for $N=1$. The parameters are set as $t_1=0.5\xi$, $\omega_e=\omega _c=20\xi$.}
\label{f2}
\end{figure}

We begin with the simplest case with $N=1$, in which the TGA is composed of one $A$ site and one $B$ site as shown in Fig.~\ref{f2} (a). An analytical expression for the reflection amplitude $r$ can be obtained as
\begin{equation}
\label{eq:7}
\begin{split}
r&=\frac{i\xi \left(\Delta_{2}+t_{1}\right)\sin k}{\left[\Delta_{1}+\xi\left(e^{ik}-1\right)\right]\left(\Delta_{2}+t_{1}\right)-J^{2}}\\
&+\frac{i\xi \left(\Delta_{2}-t_{1}\right) \sin k}{\left[\Delta_{1}+\xi\left(e^{ik}+1\right)\right]\left(\Delta_{2}-t_{1}\right)-J^{2}}-1,\\
\end{split}
\end{equation}
where $\Delta_1=E_{k}-\omega_c$ and $\Delta_2 =E_{k}-\omega_e$ represent the detuning between the propagating photon in the waveguide and the bare resonators in the CRW and the TGA, respectively. Since the spontaneous radiation for all of the resonators are ignored, it satisfies that $R+T=1$, where $R=\left|r\right|^{2}$ is the reflection rate and $T=\left|t\right|^{2}$ is the transmission rate.

In Fig.~\ref{f2} (c), we demonstrate the reflection rate $R$ as a function of the photon-atom detuning $\Delta_2$ in the resonance condition that is, $\omega_c=\omega_e$. The result for the scheme of the single cell [$N=1$ shown in Fig.~\ref{f2} (a)] is illustrated by the blue solid curve. The two complete reflection peaks ($R=1$) characterize the inter coupling inside the TGA. Besides, when only one of the two sites in the TGA couple to one resonator in the lower CRW [shown in Fig.~\ref{f2} (b)], the single photon reflection is characterized by the Rabi splitting behavior. As shown in the figure, the peaks of the orange dashed line locate exactly at the frequency $\Delta_2=\pm t_1$. The deviation from the Rabi splitting for the two sites coupling originates from the photonic interference effect when it propagates back and forth, and this interference is peculiar for the giant atom setup.

\begin{figure}[ht]
\centering
\includegraphics[width=0.89\columnwidth]{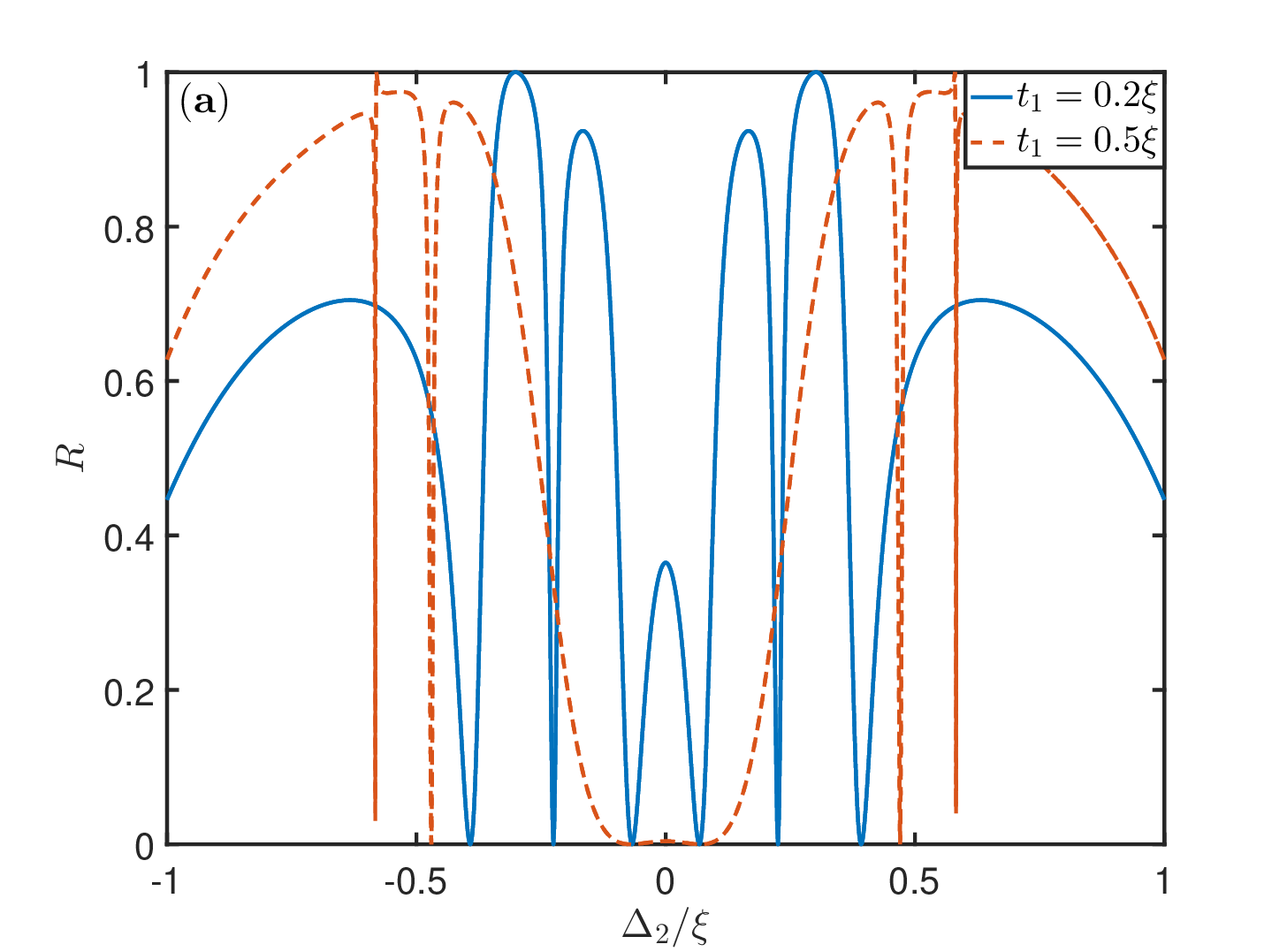}
\includegraphics[width=0.89\columnwidth]{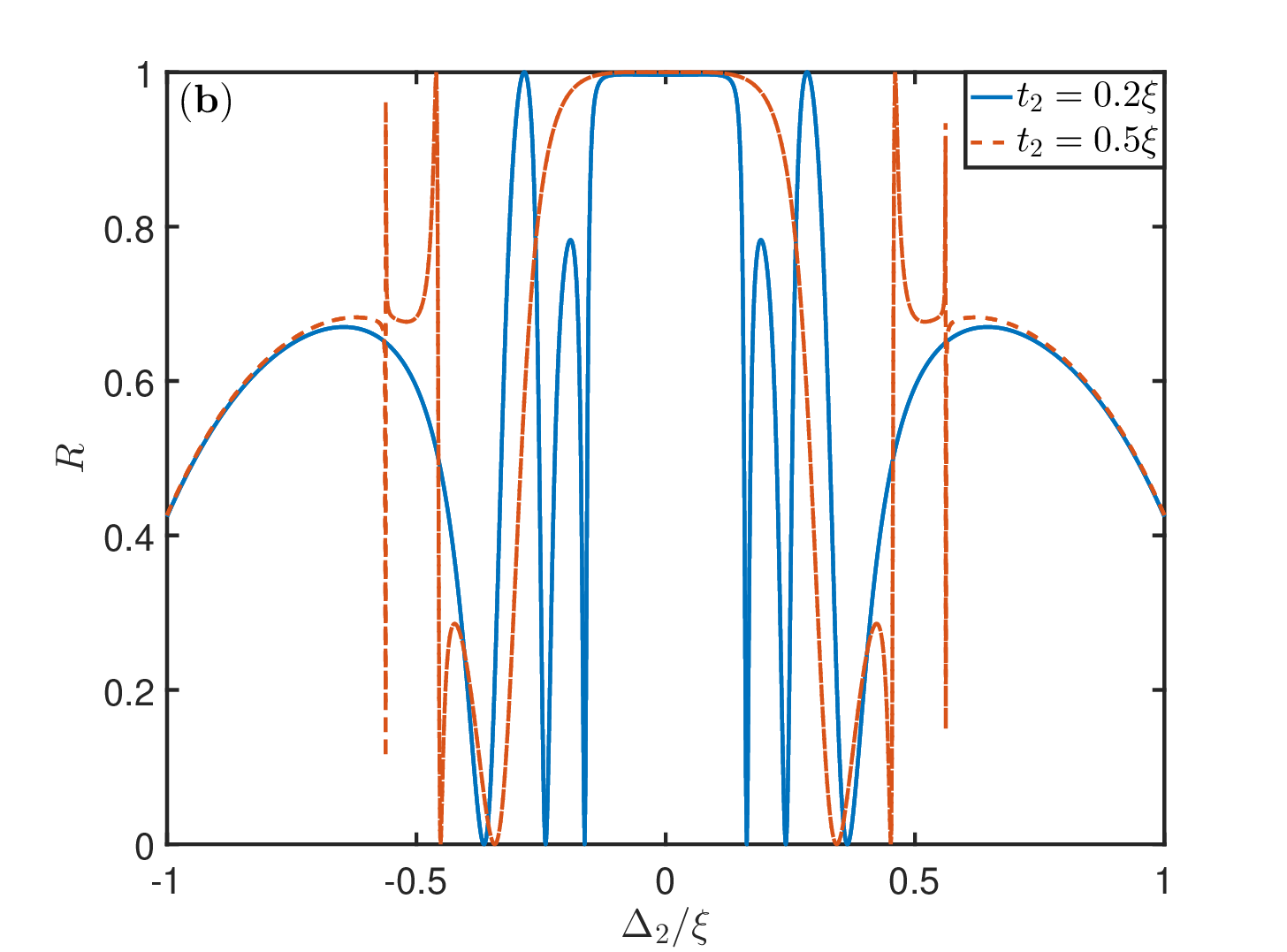}
\caption{The reflection rate $R$ as a function of detuning $\Delta_2$ for $N=5$, and (a) $t_2=0.1\xi$, (b) $t_1=0.1\xi$. The parameters are set as $J=0.9\xi$, $\omega _e=\omega _c=20\xi$.}
\label{f4}
\end{figure}

\begin{figure}[ht]
\centering
\includegraphics[width=0.89\columnwidth]{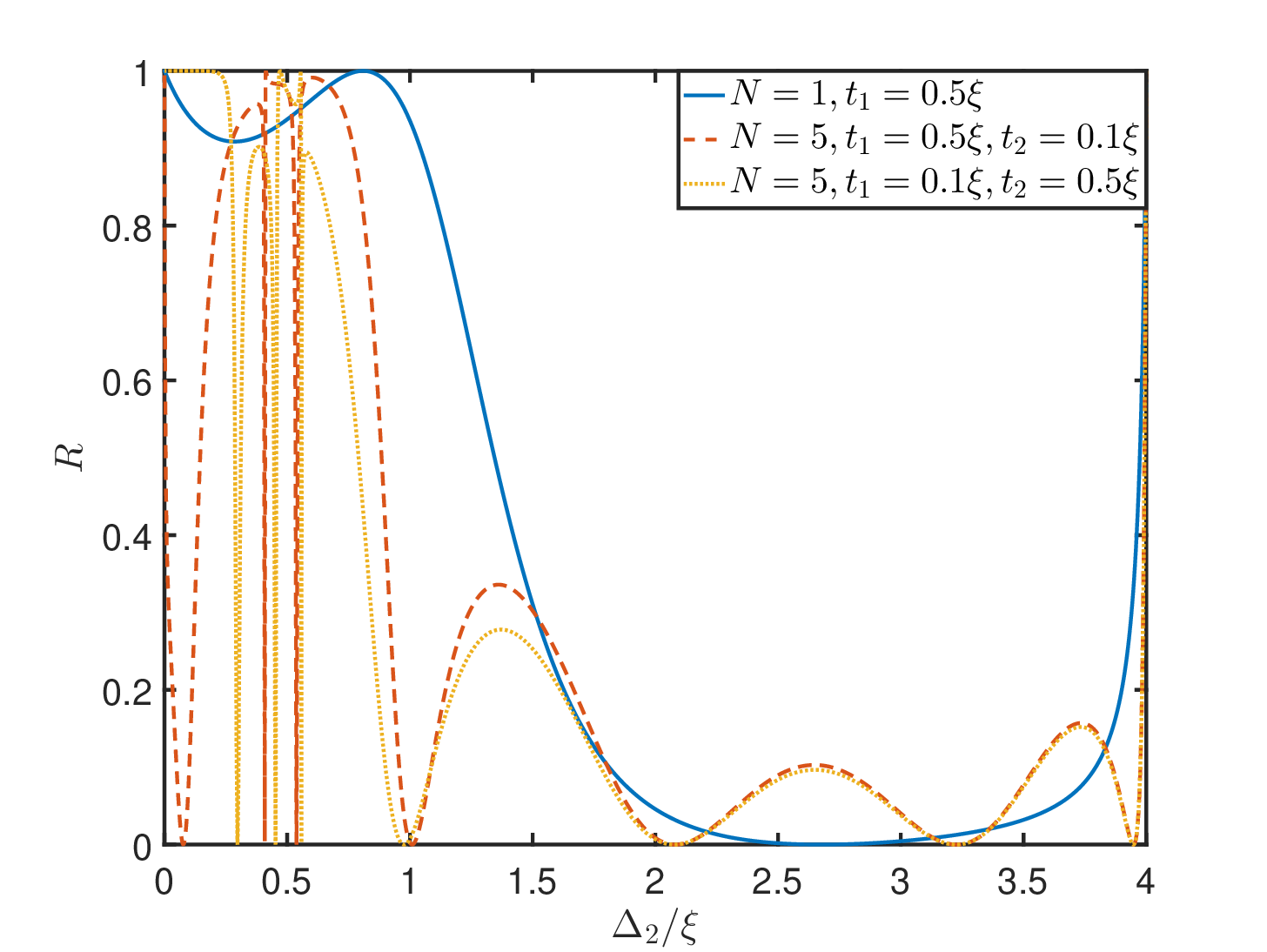}
\caption{The reflection rate $R$ as a function of detuning $\Delta_2$ when the bare resonator is off resonant with the TGA. The parameters are set as $J=0.9\xi$, $\omega _e=18\xi$, and $\omega _c=20\xi$.}
\label{fano}
\end{figure}

Next, let us move to the case of $N>1$, in which the topological properties of the TGA take effect. In Fig.~\ref{f4}, we demonstrate the reflection rate $R$ for $N=5$ as a function of the detuning $\Delta_2$. As shown in the figures, the reflection rate shows a complicated behavior in the two bands regime of $-|t_1+t_2|<\Delta_2<-|t_1-t_2|$ and $|t_1-t_2|<\Delta_2<|t_1+t_2|$. In the gapped regime with $-|t_1-t_2|<\Delta_2<|t_1-t_2|$, the reflection rate behaves differently when the TGA undergoes the topological phase transition. For example, in the topological trivial phase ($t_1>t_2$), we illustrate the results for $t_2=0.1\xi$ in the cases of $t_1=0.2\xi$ and $t_1=0.5\xi$ respectively in Fig.~\ref{f4}(a). For small $t_1$ with $t_1=0.2\xi$, there exists a small peak at the zero detuning $\Delta_2=0$. As the increase of $t_1$, the peak is smoothed and it forms a relatively wide band for $R=0$ nearby the regime of $\Delta_2=0$. It implies that the incident photon will be completely transmitted. This is due to the constructive interference caused by the photonic back and forth propagation in the spatial regime covered by the TGA. It implies that the TGA in the topological trivial phase can be used to realize quantum cloaking~\cite{fle2013,val2020}, that is, the presence of TGA has no effect on the propagation of the photon in the CRW. In other words, the photon in the CRW can not ``see'' the TGA, i.e., the TGA is invisible. On the other hand, when the TGA works in the topological non-trivial phase, the destructive interference between photons results in complete reflection and the reflection window ($R=1$) is demonstrated in Fig.~\ref{f4}(b) nearby the resonant regime of $\Delta_2=0$.

The results in Fig.~\ref{f4} indicate that, we can use the TGA to construct the coherent single-photon device. When the value of $t_{1(2)}$ be is achieved by the order of $100$ MHz experimentally~\cite{kim2021}, we can realize the quantum cloaking or complete reflection with the bandwidth of about $100-200$ MHz, which in the same order of $t_1$ and $t_2$.

We can continue to discuss the single photon scattering when the bare resonator is not in resonance with the TGA. In Fig.~\ref{fano}, we demonstrate reflectance $R$ as a function of photon-atom detuning $\Delta_2$ under non-resonant condition, i.e. $\omega_{e}\neq\omega_{c}$.
The result of the single-cell scheme ($N=1$) is illustrated by the solid blue line. The dotted orange and yellow lines represent the result in the topologically trivial and non-trivial phase respectively for $N=5$.
We find that the asymmetric reflection curve is completely different from the resonance condition, and around the reflection point, the reflection yields a Fano shape \cite{fano1961,fano2005}. With the increase of the size of the TGA, the reflection rate shows a complicated dependence on the detuning $\Delta_2$. This means that off resonance can induce Fano physics in such a hybrid system.
We also observe that, there will be one or more completely reflection frequencies (excluding the edge of the photonic propagation band), depending on the values of $N$. It implies that, we can design on-demand single-photon transistors by adjusting the size of TGA in our waveguide QED setup.

\section{Single photon bound state}
\label{BOUND}

In the above section, we have studied the single photon scattering states which locate inside the propagating band $E_{k}\in\left[\omega_{c}-2\xi,\omega_{c}+2\xi\right]$. Due to the coupling between the TGA and the CRW, the translational symmetry of the system is broken, leading to the photonic bound states which lay outside the propagating band. Here, we resort to the numerical diagonalization of the Hamiltonian to find these bound states when the TGA works in the topologically non-trivial phase ($t_1<t_2$).

\begin{figure}
\begin{centering}
\includegraphics[width=\columnwidth]{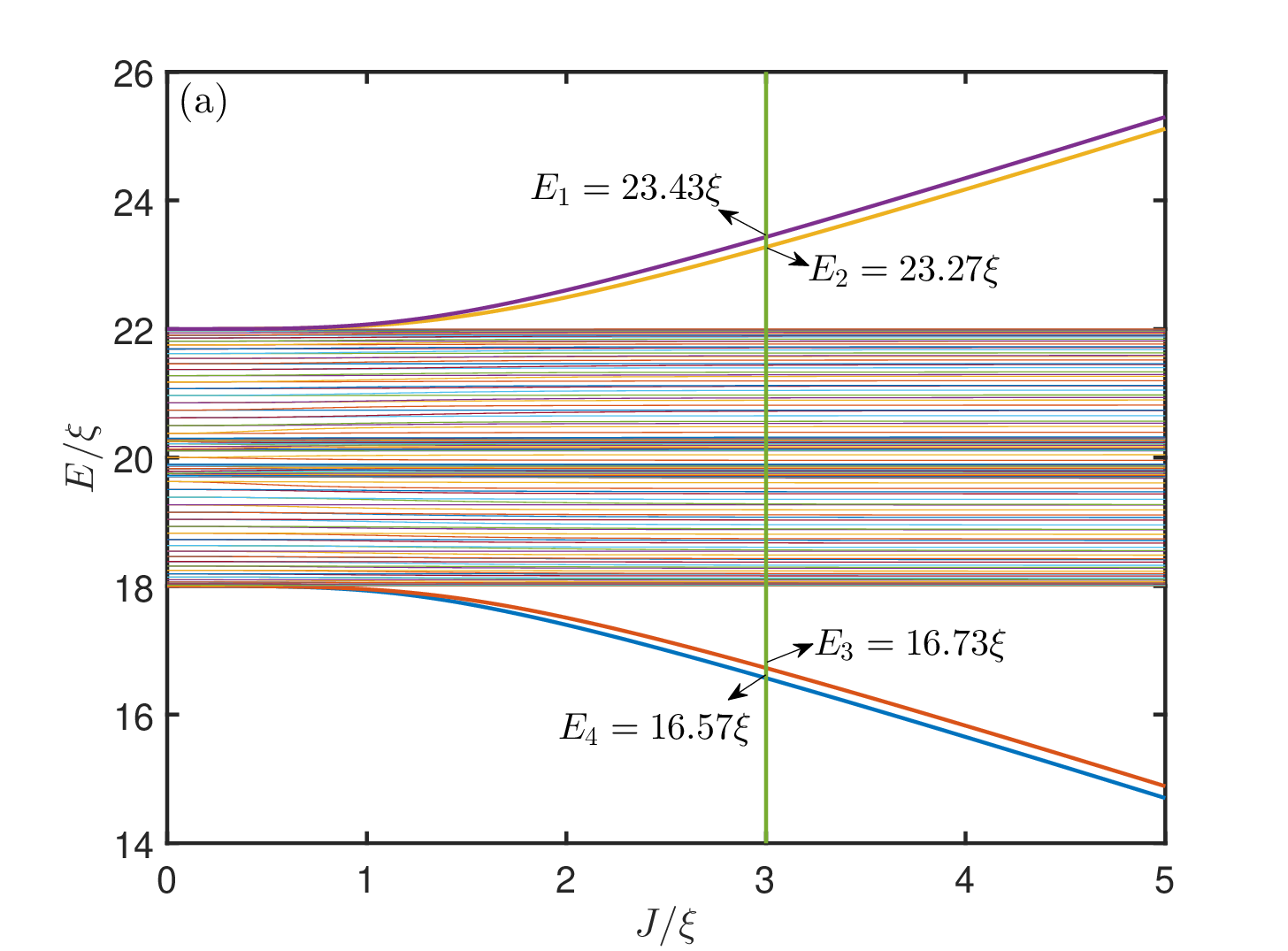}
\includegraphics[width=\columnwidth]{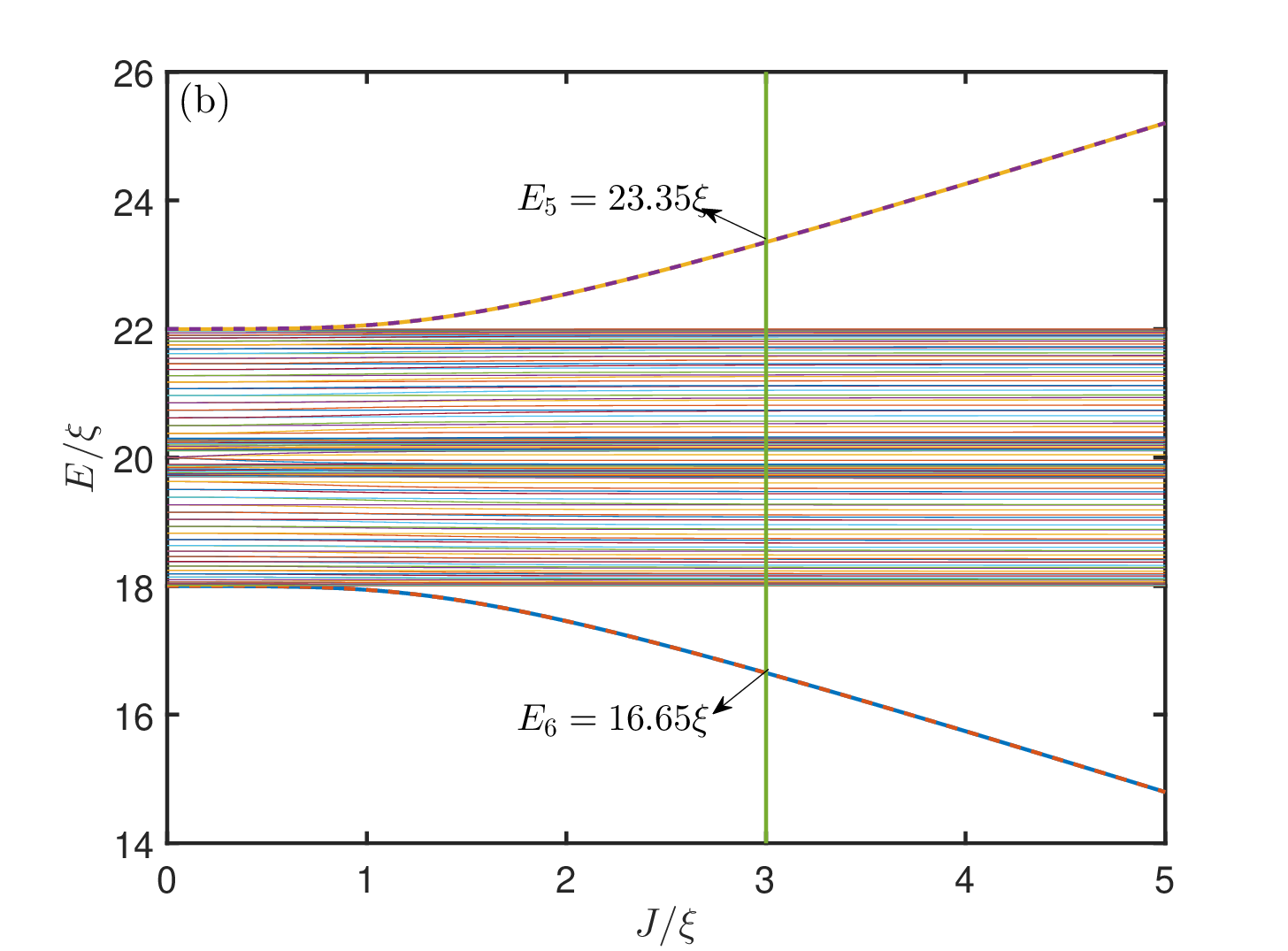}
 \par\end{centering}
 \caption{The energy spectrum for $N=29$ in (a) $t_3=t_2$, and (b) $t_3=0$. The parameters are set as $t_1=0.1\xi$, $t_2=0.2\xi$, $\omega _e=\omega _c=20\xi$.}
  \label{f5}
\end{figure}

In Figs.~\ref{f5} (a) and (b), we plot the energy spectrum by considering that the TGA works in the PBC $(t_{3}=t_{2})$ and OBC $(t_{3}=0)$, respectively. As expected, we observe the continual band which supports the propagating modes in both of the cases and the difference arises from the bound states outside the continuum. In the PBC, there exist two pairs of non-degenerate bound states, which locate symmetrically above and below the continuum and are denoted by $E_1, E_2,E_3$ and $E_4$, respectively as shown in Fig.~\ref{f5} (a). Within the available experimental parameter $\xi=100$\ MHz, the gap between the gap state $\Delta_E=E_1-E_2=E_3-E_4$ is in the order of $10$\ MHz in the condition of $J=3\xi$. As a comparison, when the TGA works in the OBC, the pairs of the bound states become degenerate as shown in Fig.~\ref{f5} (b). Moreover, it satisfies $E_5=(E_1+E_2)/2$ and $E_6=(E_3+E_4)/2$.

\begin{figure}
\begin{centering}
\includegraphics[width=\columnwidth]{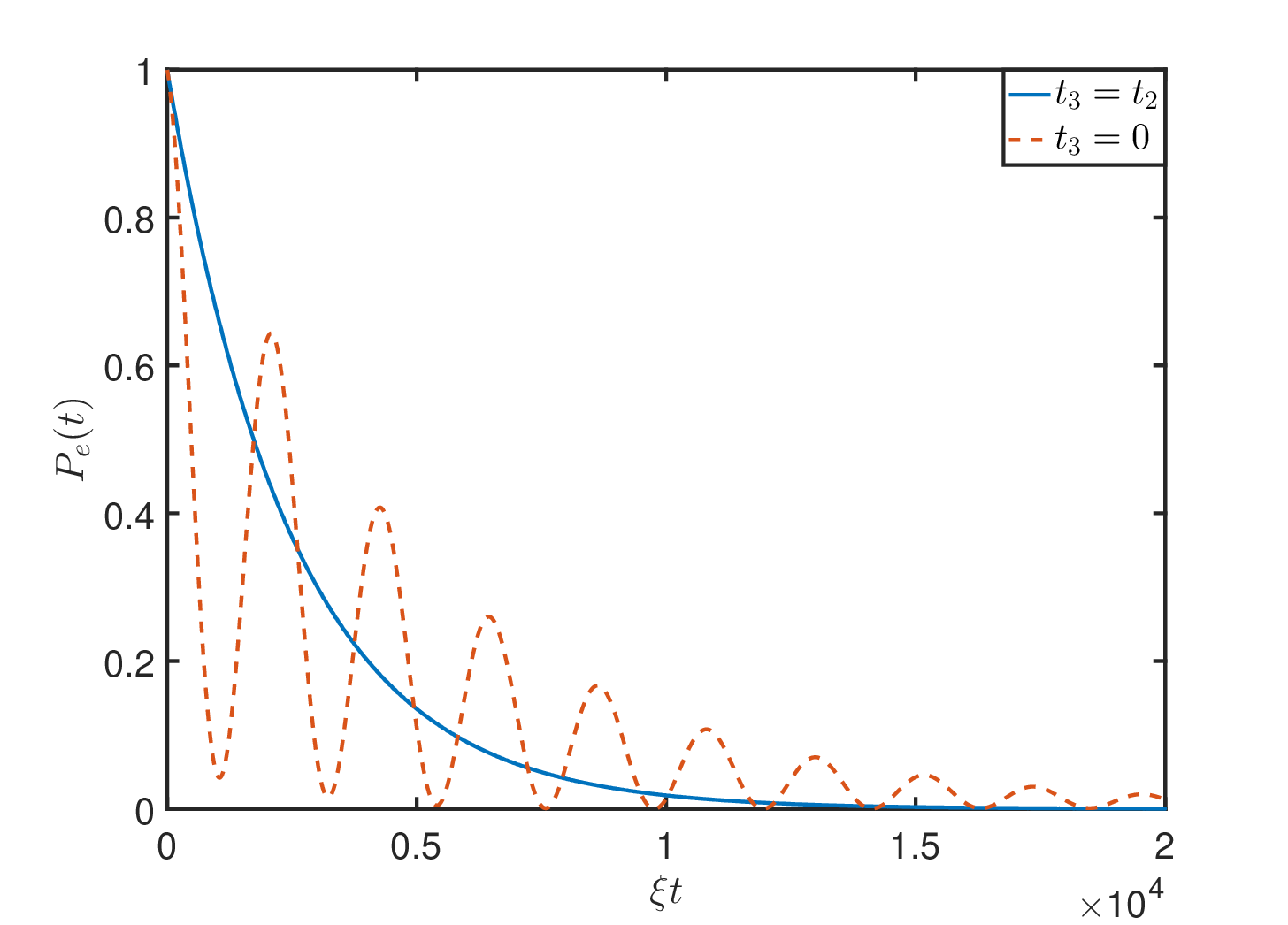}
 \par\end{centering}
 \caption{The evolution of the population in the excited state for the probing atom for $t_2=t_3$ (blue solid line), and $t_3=0$ (orange dashed line). The parameters are set as $N=29$, $\omega _e=\omega _c=20\xi$, $\omega_p=E_6=16.65\xi$, $t_1=0.1\xi$, $t_2=0.2\xi$, $J=3\xi$, $f=2\times10^{-3}\xi$, $\gamma=2\times10^{-4}\xi$.}
  \label{f7}
\end{figure}

The above transition from the degenerate bound states to non-degenerate ones provides us an effective approach to detect the boundary condition of the TGA. To this end, we introduce an auxiliary probing atom whose resonant frequency satisfies $\omega_p=E_6$ and observe its excitation evolution. Then, the probing Hamiltonian is expressed as
\begin{equation}
H_p=H+(\omega_p-i\gamma)\left|e\right\rangle _{p}\left\langle e\right|+f\left(\tau_{+}a_{j}+{\rm H.c.}\right)
\end{equation}
which demonstrates that the probing atom is coupled to the $j$th cavity with the coupling strength $f$, $\gamma$ and $\tau_{+}=\left|e\right\rangle _{p}\left\langle g\right|$ are the spontaneous emission and raising operator of the probing atom.

Preparing the probing atom in the excited state and the TGA-CRW coupled system in the ground state initially, we plot the excitation dynamics of the probing atom in Fig.~\ref{f7} by taking $J=3\xi$. Here the probing atom is located in the resonator of $j=0$, that is, the left leg of the TGA.
Since the probing atom is large detuned from both of the bound states in the condition of $f\ll|\omega-E_{3(4)}|$, the probing atom is effectively decoupled from the TGA-CRW system in the PBC. As a result, the dynamics of $P_e(t)=\langle |e\rangle_p\langle e|\rangle$ shows an exponential decay, which is determined by the spontaneous dissipation rate $\gamma$ as shown by the solid line in Fig.~\ref{f7}.
In the OBC, the pair of non-degenerate bound states emerge into the degenerate ones, which resonate with the probing atom. So that, the dynamics exhibits a Rabi oscillation character as shown by the dashed line. In this sense, the boundary condition of the TGA can be detected in a coherent manner.

\section{REMARKS and CONCLUSIONS}

\label{conclusion}

In this paper, we have constructed a TGA via a finite SSH chain which couples to the CRW. In the single excitation subspace, we discuss the scattering and bound states, respectively. We find that the photonic scattering behavior can be modulated by the topological nature of the TGA. Together with the interference effect during the propagation of the photon in the waveguide, we can design the wide band cloaking or reflection photonic device when the TGA works in the topological trivial or non-trivial phases, respectively. As for the bound states which locate outside the continuum, the energy gap can be closed or opened, depending on the boundary condition of the TGA.

In the current experimental availability, the proposed model can be realized in the superconducting circuits. In 2014, the first experiment which couples the superconducting transmon quantum qubit and the surface acoustic waves (SAWs) waveguide was realized. Here, the qubit is $20$ times larger than the wavelength of the SAWs in size and it therefore reached the giant atom regime~\cite{gus2014}. This ratio between the size of the giant atom and the wavelength was then developed to $100$ and the light-matter interaction was achieved by tens of MHz~\cite{andg2019}.
Additionally, the giant atom was also achieved by coupling artificial atoms created with Josephson junctions to superconducting circuits through the capacitance or inductance. In Ref.~\cite{Kanb2020}, the giant atom was experimentally coupled to a waveguide at multiple, yet well separated locations. The distance between two coupling points could reach $20.54$ mm~\cite{vada2021}.
Recently, the topology of two nested giant spin ensembles (GSE) has been experimentally demonstrated, the distance between the two inner (outer) coupling points of the is designed to be $8.3\,(16.6)$cm~\cite{wangzq2022}. It provides a new platform for ``giant atom" physics.
Moreover,the CRW constructed with $9$ superconducting qubits is also realized in the experiment, where the nearest-neighbor coupling strength is $\xi/2\pi=50$ MHz~\cite{rou2017}.
In a recent experiment, Painter's group expanded the CRW consist of a $42$ unit-cell array of capacitively coupled lumped-element microwave resonators~\cite{zhangxy2023}, and the SSH topological structure was also constructed with the hopping strength $J(1\pm\delta)$ where $J/2\pi=368$ MHz, $\delta=0.282$. Furthermore, the qubit-resonator coupling strength in such structure was achieved by $g/2\pi=124.6$ MHz~\cite{kim2021}

The establishment of TGA in our work provides an unconventional approach to regulate the transmission and distribution of the single photon in the CRW. Here, the interference effect for the propagation of the photon in the regime of the giant atom will be applicable in quantum control and quantum information processing.
We hope that our work of controlling photons through the giant atom systems combined with topologies will stimulate further research of the hybrid system and broaden the range of application of artificial giant atom.

\begin{acknowledgments}
This work is supported by the funding from Jilin Province (Grant Nos. 20230101357JC and 20220502002GH) and the National Science Foundation of China (Grants No. 12105026 and No. 12375010).
\end{acknowledgments}

\end{document}